# Two-dimensional atomic-scale ultrathin lateral heterostructures

Nanami Ichinose[1], Mina Maruyama[5]*, Takato Hotta[1], Zheng Liu[3], Ruben Canton-Vitoria[1], Susumu Okada[5], Fanyu Zeng[1], Feng Zhang[1], Takashi Taniguchi[3], Kenji Watanabe[4], and Ryo Kitaura[3,*]

[1] *Department of Chemistry, Nagoya University, Nagoya, Aichi 464-8602 Japan*

[2] *Innovative Functional Materials Research Institute, National Institute of Advanced Industrial Science and Technology (AIST), Nagoya, Aichi 463-8560, Japan*

[3] *Research Center for Materials Nanoarchitectonics, National Institute for Materials Science, 1-1 Namiki, Tsukuba 305-0044, Japan*

[4] *Research Center for Functional Materials, National Institute for Materials Science, 1-1 Namiki, Tsukuba 305-0044, Japan*

[5] *Department of Physics, Graduate School of Pure and Applied Sciences, University of Tsukuba, 1-1-1 Tennodai, Tsukuba, Tsukuba 305-8571, Japan*

Corresponding Author: R. Kitaura, KITAURA.Ryo@nims.go.jp, r.kitaura@nagoya-u.jp; M. Maruyama, mmaruyama@comas-tsukuba.jp

ABSTRACT

Ultrathin lateral heterostructures of monolayer $MoS_2$ and $WS_2$ have successfully been realized with the metal-organic chemical vapor deposition method. Atomic-resolution HAADF-STEM observations have revealed that the junction widths of lateral heterostructures range from several nanometers to single-atom thickness, the thinnest heterojunction in theory. The interfaces are atomically flat with minimal mixing between $MoS_2$ and $WS_2$, originating from rapid and abrupt switching of the source supply. Due to their peculiar one-dimensional/two-dimensional mixed-dimensional structures, the resulting ultrathin lateral heterostructures can show emergent properties. Indeed, we have theoretically revealed that the spatial dimensionality of carriers in nanoscale lateral $MoS_2/WS_2$ can be tuned by gate voltages: trans-dimensional materials. The MOCVD growth developed in this work allows us to access various ultrathin lateral heterostructures, leading to future exploration of their emergent properties absent in each component alone.

The spatial dimension plays an essential role in materials science. When the dimension of a system is reduced, electronic band structures can be drastically altered, leading to the emergence of peculiar optical and electronic properties. For example, thinning graphite down to its two-dimensional (2D) counterpart, namely graphene, leads to the emergence of Dirac cones around the fermi level, where massless Dirac fermions dominate the optical and electronic properties.[1-4] Another remarkable example is transition metal dichalcogenides (TMDs). When TMDs, such as $MoS_2$ and $WS_2$, are thinned down to monolayer, the valley degree of freedom (VDF) emerges due to the inversion symmetry breaking.[5] Also, distinct from multilayers, monolayer TMDs are direct gap semiconductors, providing the opportunity to explore exciton physics and control VDF by circularly polarized light excitations.[6-10] As shown in these examples, reducing the spatial dimension of materials can significantly impact their electronic band structures and physical properties.

The exploration of 2D systems has been one of the central issues in materials science for the past couple of decades. Although high-quality 2D systems were limited to semiconductor quantum wells (QWs) in the past, the emergence of graphene in 2004 has drastically altered the situation. The discovery of graphene has not only led to the exploration of the exotic physics of graphene and its potential for a variety of applications but also stimulated the search for other 2D systems, providing various 2D materials, including 2D semiconductors, metals, topological insulators, and ferromagnets.[11-13] Furthermore, these 2D-materials families can be readily prepared through the simple mechanical exfoliation technique, giving rise to a significant widening of the scope of 2D-materials research.

2D materials, moreover, provide a fascinating opportunity to explore a new direction, e.g., one-two-mixed-dimensional (1-2mD) systems.[14] The stitching of 2D materials side-by-side leads to forming a one-dimensional (1D) version of QWs, where 1D-stripe wells align to form 1-2mD structures. These are 2D structures with monolayer thickness, but their electronic structure can possess 1D nature depending on the stripe width. Because of the unique structure, 1-2mD structures (1D QWs) can have various applications for optoelectronic devices like conventional QWs based on bulk semiconductors.[15,16] Also, 1-2mD structures possess 1D interfaces, which can host peculiar 1D edge/interface states, and spatial symmetry different from their components alone. Furthermore, when the width of 1D wells becomes nanoscale, the enhanced many-body effect is expected to lead to novel phenomena. 1-2mD structures, therefore, can provide a fascinating platform to explore physics and applications in the realm of low-dimensional materials.

However, controlling the width of 1-2mD structures down to the nanometer scale has technically been challenging. Pioneering work by Ajayan's group has shown that lateral heterostructures can be grown by the chemical vapor deposition (CVD) method.[17] Quite a few papers, then, have reported CVD growth of lateral heterostructures, demonstrating the formation of 1-2mD structures with well-width ranging typically from several tens of nanometers to micrometers.[18-20] Metal-organic CVD (MOCVD) has also been applied to grow 1-2mD structures, forming multiple lateral heterostructures with a well-width of several tens of nanometers.[21,22] Ultrathin 1D wells, whose width is subnanometer ~ several nanometers, have recently been reported based on a different approach, the dislocation-based method.[23,24] In this approach, dislocations at the interface of lateral heterostructures can propagate, leading to the formation of ultrathin 1D QWs. Although significant advancement in the realization of 1-2mD structures has been achieved so far, controllable growth of nanometer-scale 1-2mD structures is still quite challenging.

Here we report the MOCVD growth of subnanometer-scale 1-2mD structures composed of $MoS_2$ and $WS_2$. We have selected MOCVD for the growth of $MoS_2$-$WS_2$ 1-2mD structures because rapid switching of gaseous organometallic sources facilitates the synthesis of 1-2mD structures with atomic precision. To realize multiple abrupt switching of the source supply, we have developed an MOCVD setup, where organometallic compounds and organic sulfide can be supplied through automatic valve control. Slow growth rate and swift valve operation can thin the width of 1D-stripe wells down to single-atomic width. In addition, atomic-resolution scanning transmission electron microscopy (STEM) has successfully visualized atomically sharp interfaces in 1-2mD structures. Density functional theory (DFT) calculations have revealed that the ultrathin 1-2mD systems possess a direct bandgap with type-II band alignment, and their dimensionality can be tubed by fermi-level control. These findings have clearly demonstrated that 1-2mD structures with atomically sharp junctions can possess properties drastically different from each component, giving us a versatile platform to explore novel phenomena from the mix-dimensionality.

Figure 1(a) shows a schematic representation of the MOCVD setup (see also a photograph of the setup in Fig. S1). This setup is a vertical-type cold-wall MOCVD, where gaseous sources are supplied through a showerhead right above a growth substrate; the substrate-showerhead distance can be adjusted with a lifting mechanism. The chamber wall and the showerhead are cooled by water circulation, which prevents materials deposited on the wall from desorption during growth processes. The source-supply lines are separated from the chamber by solenoid valves, which can be operated quickly (response time ~ millisecond) and automatically by a valve-control system. Because the valves for switching the sources are right above the chamber, the dead volume for source supply is minimal, and thus the valve operations can swiftly change the source supply during growth processes. This rapid switching of source supply is a significant advantage for the growth of 1-2mD structures against the conventional solid-source-based CVD method, where solid sources cannot be exchanged instantaneously.[25] We investigated the precursor flow in the MOCVD chamber by the finite element method (FEM). Figure 1(b) shows the source flow and the normalized precursor concentration distributions. The concentration distribution on the substrate surface shows good uniformity with a slight difference (8 %) between the center and corners. The uniformity probably results from the geometry of our setup, where sources are mixed before going through the showerhead.

For MOCVD growth of 1-2mD structures, volatile compounds, including diethylesulfide, bis(*ter*-butylimido)-bis(dimethylamido)tungsten(VI) ($W(TBI)_2(DMA)_2$), and $Mo(TBI)_2(DMA)_2$, were employed as growth precursors. These are volatile liquids at room temperature and can be readily supplied via bubbling the liquid sources with a flow of Ar to the growth chamber, where supplied sources thermally decompose to form TMDs flakes on growth substrates. Figure 2(a) shows optical microscope images of $WS_2$ formed on a hexagonal boron nitride (hBN) flake on a soda-lime glass substrate. Although contrast in the optical microscope image is weak, we can see clear triangular contrast in the corresponding PL image (Fig. 2(b)), consistent with single crystals of monolayer $WS_2$. We used soda-lime glass substrates because sodium can be directly supplied during growth processes, acting as a growth promoter to form monolayer $WS_2$ on hBN.[26,27] $WS_2$ grown on hBN possess the atomically flat structure, eliminating the substrate effect to observe PL/Raman spectra with sharp peaks and AFM images with minimal root mean square roughness.[28,29] A typical Raman spectrum of $WS_2$ grown on hBN shows Raman bands from $A'_1$ and 2LA mode at 420 and 357 $cm^{-1}$, respectively (Fig. 2(c)). A corresponding PL spectrum shows the pronounced peak from the radiative recombinations of K-K direct excitons (Fig. 2(d)). A typical AFM image and the corresponding height profile demonstrate that grown flakes are monolayer, consistent with the Raman and PL spectrum (Fig. 2(e)).[30] It should be noted that the G and D band (1300~1600 $cm^{-1}$), as well as methyl groups (2800-3200 $cm^{-1}$), are absent in Raman spectra of grown $MoS_2$ and $WS_2$, indicating that carbonaceous species arising from the decomposition of $W(TBI)_2(DMA)_2$, $Mo(TBI)_2(DMA)_2$ and diethylesulfide are not incorporated in the final products (Figs. S2). The successful formation of $WS_2$ crystal means that organic functional groups in the precursors are effectively removed and not included in $WS_2$ in the growth process. $MoS_2$ can be grown under a similar growth condition (see Figs. S3).

In addition to the rapid switching of source supply, the control of nucleation processes is crucial for the growth of 1-2mD structures; the growth of crystals side-by-side with minimal additional nucleation is essential. Because nucleation is driven by the degree of supersaturation, we control the nucleation rate by variating the supply rate of the metal precursors. The supply of the diethylesulfide has been kept constant throughout growth processes; we used sulfur/metal molar ratios larger than $10^4$ to minimize the formation of sulfur vacancies. Figure 3(a) shows a typical growth process for $MoS_2/WS_2$ 1-2mD structures. Before the first growth, a substrate was annealed at growth temperature under a flow of diethylesulfide to clean the surface, and then nucleation of the first $WS_2$ was performed with a relatively high nucleation rate to have monolayer $WS_2$ for the following side-by-side growth. The source supply was then alternated between $Mo(TBI)_2(DMA)_2$ and $W(TBI)_2(DMA)_2$ with a supply rate low enough to minimize additional nucleations, leading to consecutive growth of $MoS_2/WS_2$ at the edges of the existing layers. Detailed conditions are given in the caption of Fig. 3(a).

Figure 3(b) shows an AFM phase image of a $MoS_2/WS_2$ 1-2mD structure grown on an hBN flake; this growth experiment used 60-second Mo and W source supplies. As seen in the AFM images, a periodic stripe contrast exists in the AFM image. The height profile shows the height of these samples is ~ 0.7 nm, consistent with the monolayer structure. Although the difference in the height of $MoS_2$ and $WS_2$ is tiny, the ultraflat structure of the 1-2mD sample grown on hBN enables us to observe the 1D stripe structures directly in the AFM images. Direct inspection of the AFM image gives the stripe width of about 10 nm, indicating that we can control the width of the 1D stripe down to the sub-10-nanometer range by the current MOCVD growth.

For precise structural characterization, including stripe width and interface, of sub-10-nanometer 1-2mD structures, we have performed high-angle-annular-dark-filed STEM (HAADF-STEM) observations of a sample prepared with 60-second Mo and W source supplies. Figure 4(a) shows a high-magnification HAADF-STEM image of a $MoS_2/WS_2$ 1-2mD structure. The image shows the atomic structure of the 1-2mD, where the hexagonal network structure composed of metal and chalcogen atoms is clearly visualized. The spot-like contrasts can be categorized into the weakest, middle, and strongest. In the HAADF-STEM images, the image contrast significantly depends on atomic number; the contrast becomes stronger as the atomic number increases (Z-contrast imaging). The weakest one, therefore, corresponds to S atoms, whereas the middle and strongest ones correspond to Mo and W atoms, respectively. To confirm this assignment, we have performed an image simulation with the structure model shown in Fig. 4(b).[31] As seen in Fig. 4(a), the simulated image reproduces the observed one well, confirming the assignment of image contrasts. Also, energy-dispersive X-ray spectroscopy has revealed that W (Mo) localizes at bright (dark) contrast regions, whereas S uniformly distributes over the sample (Fig. S4). These electron microscopy results are consistent with a 1-2mD structure composed of monolayer $MoS_2$ and $WS_2$.

The width of the 1D-stripe $MoS_2$ shown in Fig. 4(a) is only 3 nm, one of the smallest among TMD-based lateral heterostructures ever reported. The typical growth rate used for 1-2mD structures is several nanometers/minute, and therefore, valve operation with a duration time of 60 seconds can yield this ultra-narrow 1D stripe of $MoS_2$. The interface between $WS_2$ and $MoS_2$ is parallel to the zig-zag direction with an atomically sharp junction structure, different from those formed via the dislocation-driven approach.[23] Because of anisotropy of interfacial free energy, straight zig-zag edges preferentially appear under the present growth condition (facet growth), moving forward by incorporating metal chalcogen atoms to form the atomically sharp interfaces along the zig-zag direction.

The growth temperature of ~ 1000 K ($k_BT$ ~ 86 meV) is much lower than the bonding energy between Mo/W and S (the energy scale of ~ 1eV), and thermal roughening, which can cause roughening of $MoS_2/WS_2$ interfaces, should not occur in the current MOCVD growth processes. On the other hand, unstable atoms attached to edges can diffuse along the edges at the growth temperature, leading to structural reconstruction to form atomically straight edges for the 1-2mD structure observed in the HAADF-STEM image. The sharp interfaces also mean that diffusion of Mo (W) to the $WS_2$ ($MoS_2$) region does not occur once interfaces are formed, consistent with stable chemical bonds between Mo/W and S. The amount of W (Mo) contaminated in the $MoS_2$ ($WS_2$) region is typically less than 1 % (see Fig. S5), which is significantly smaller than those reported previously. Because we put valves right above the main chamber to minimize dead volume, the valve operation can quickly switch source supply to reduce the incorporation of W (Mo) at $MoS_2$ ($WS_2$) regions.

As demonstrated, the newly developed MOCVD setup can realize nanoscale stripe width through a few tens of seconds of valve operations. This finding motivated us to realize atomic-scale junction widths by reducing valve switching time to several seconds. Figures 5(a) and (b) are HAADF-STEM images of 1D stripes of $MoS_2$ grown with shorter switching times of 15 – 30 seconds. As you can see, the resulting stripe widths of $MoS_2$ are 1.5 and 0.8 nm, subnanometer atomic-scale stripes. We also found that the MOCVD setup can yield single atomic stripes, ultimately thin junctions; contrasts arising from single-atomic linear arrangements of Mo (W) in $WS_2$ ($MoS_2$) are visualized in Fig. 5(c). These images demonstrate that the structural control with the present MOCVD method can reach the atomic level, and thus the current method can be a versatile tool in the future exploration of the properties of tailor-made 1-2mD structures.

Density functional theory (DFT) calculations have revealed their peculiar electronic band structures. Figures 6 (a) and (b) show the band structure and local density of states of a $WS_2/MoS_2$ lateral heterostructure, $((WS_2)_{10}/(MoS_2)_{10})_n$, whose stripe width is 2.74 nm (the unit cell and the corresponding Brillouin zone are shown in Figs. S6). As seen in Fig. 6(a), the $WS_2/MoS_2$ possesses a direct bandgap, unlike vertical heterostructures, whereas its edge band alignment is type II, as is the vertical heterostructures; similar trends can be seen in $WS_2/MoS_2$ with different widths (Fig. S7). Electronic states mixing between $WS_2$ and $MoS_2$ is not significant, and the valence band maximum (VBM) and the conduction band minimum (CBM) distribute in $WS_2$ and $MoS_2$ parts, respectively (Fig. 6(b), wavefunctions at CBM and VBM are shown in Fig. S8(a)). Because of the localization nature, the electronic band structure of the $MoS_2/WS_2$ is expected to show an anisotropic 1D character. The effective masses of electrons ($m_e$) and holes ($m_h$) at CBM and VBM along the X-Γ line are 0.46 $m_0$ and 0.51 $m_0$, respectively. Whereas band dispersion along Γ-Y line is almost flat, clearly demonstrating the 1D nature of the electronic band structure of $MoS_2/WS_2$.

The band offset at CBM is about 0.4 eV, which should act as a 1D confinement energy barrier for electrons, as schematically shown in lower left of Fig. 6(c). For visualizing the 1D confinement, electron injection into lateral heterostructures has been theoretically investigated with the dual-gate field-effect transistor structure shown in upper left of Fig. 6(c); an effective screening medium method is employed to simulate the top and bottom gate electrodes with infinite permittivity[32]. As you can see in the isosurface of electron density (Fig. 6(c) right), electrons mainly distribute in the $MoS_2$ region under an electron density of $10^{12}/cm^2$, emerging 1D channels in a 2D structure. Electron density penetrates the $WS_2$ region due to the penetration of the $MoS_2$ CBM wavefunction with a penetration length of 1.6 nm (see Fig. S8). When electron density increases to $10^{13}/cm^2$, electrons start to spill over the $WS_2$ more and show a more 2D nature. In this calculation, we use $((WS_2)_{10}/(MoS_2)_{10})_n$, whose lateral periodicity is 5.48 nm, due to computational limitations. $WS_2$ width can be much broader in actual samples, where smaller electron densities, so as small gate voltages, can induce crossover from 1D electrons at $MoS_2$ stripes to 2D electrons over $MoS_2$ stripes and 2D $WS_2$; charging energy at 1D stripes is also expected to facilitate the 1D to 2D crossover (supplemental discussion 1 gives a related discussion). This result clearly demonstrates that we can control the dimensionality of electrons in the 1-2mD structures of $WS_2/MoS_2$ by simply applying gate voltages. The same gate-induced dimensionality alternation should be seen in other 1-2mD structures with relatively small band offset, although the dimensionality of carriers is usually characteristic of each material. The trans-dimensionality is a unique characteristic of 1-2mD structures, providing a playground for exploring emergent gate-induced properties in the future.

In conclusion, we have successfully demonstrated the growth of 1-2mD structures composed of $MoS_2$ and $WS_2$ by the MOCVD method. Atomic-resolution HAADF-STEM observations have revealed that the junction widths of 1-2mD structures range from several nanometers to single-atom thicknesses. The interfaces are atomically flat with minimal mixing between $MoS_2$ and $WS_2$, originating from rapid and abrupt switching of the source supply. The resulting 1-2mD structures, ultrathin lateral heterostructures, possess spatial symmetry different from $MoS_2$ and $WS_2$ alone and electronic bandstructure, which can host peculiar interface/edge states. Also, we have revealed that carrier dimensionality can be tuned by applying gate voltage in a 1-2mD structure, which can act as a trans-dimensional system. 1-2mD structures, therefore, can have emergent optical/electronic properties, which are absent in each component. The MOCVD setup developed in this work gives a powerful tool for exploring there emergent properties of 1-2mD structures in the future.

## Methods

**Flow simulation.** The numerical model of the MOCVD chamber was developed using the commercial finite element software COMSOL, in which multi-physical fields, including laminar flow, heat transfer, and mass transfer, were coupled. Depending on the importance of each region, step-up mesh settings from 0.88 to 20 mm are used. Since the detailed reaction process of MOCVD is currently complex and unknown, the chemical reaction was not considered, and the results were analyzed using the precursor concentration as an indicator. More details about the simulation are described in supporting information.

**Ab-initio calculations**. The geometrical and electronic structures were investigated using the STATE package based on DFT. A generalized gradient approximation using the Perdew-Burke-Ernzerhof functional form was employed to describe the exchange-correlation potential energy. Electron-ion interactions were treated in terms of ultrasoft pseudopotentials. The valence wave function and deficit charge density were expanded by plane-wave basis sets with cutoff energies of 25 Ry and 225 Ry, respectively. Atomic coordinates were optimized until the forces acting on each atom were less than 5 mRy/A under the lattice parameter corresponding to the experimental value.

**MOCVD growth.** We purchased sources (diethylesulfide, bis(*ter*-butylimido)-bis(dimethylamido)tungsten(VI), and bis(*ter*-butylimido)-bis(dimethylamido)Molybdene(VI)) from Gas-Phase Growth LTD., and these sources were used as received. We performed MOCVD growth using a home-built cold-wall type MOCVD chamber. First, we prepared hBN flakes on a soda-lime glass substrate by the mechanical exfoliation method. After annealing the hBN/glass substrate under dry-air flow at 700 degrees, the substrate on a SiC susceptor was transferred from the load-lock chamber to the main growth chamber. And then, we started to flow Ar gas with a flow rate of 500 ~ 750 sccm and heat the susceptor to a reaction temperature of 700~750 degrees. After the susceptor temperature became stable, we started the supply of sources through the showerhead to the substrate. After a particular reaction time, the supply of the metal source was stopped while the supply of the sulfur source was maintained until the susceptor temperature was lower than 350 degrees. All operations, including control of temperature, flow rate, and valve open/close, were performed by a LabVIEW-based automatic control system.

**Characterizations.** Raman and PL spectra were measured with microspectroscopy systems (Jovin-Yvon HR-800 and Renishaw InVia Raman) with 488 or 532 nm CW laser excitation. The laser was focused onto samples with objective lenses (×50–100 and 0.7–0.85 NA), and Raman and PL signals were detected with a charge-coupled device (CCD). All measurements were carried out at room temperature under atmospheric conditions. PL spectra at low temperatures were obtained with a homemade microspectroscopy system. HAADF- and ABF-STEM images were taken at room temperature using a JEM-ARM200F ACCELARM (cold field emission gun) equipped with a CEOS ASCOR corrector, operated at 120 kV. For each frame of STEM image, fast scan rates of 3~ 5 μs per pixel were used. After drift compensation, several frames up to ten are overlapped to form a HAADF or ABF STEM image. EDS elemental mapping is performed by using JED-2300.

## Data availability

The data that support the plots within this paper and findings of this study are available from the corresponding author upon request.

## Acknowledgements

R.K. was supported by JSPS KAKENHI Grant No. JP23H05469, JP22H05458, JP21K18930 and JP20H05664, and JST CREST Grant No. JPMJCR16F3, SCICORP Grant No. JPMJSC2110 and PRESTO Grant No. JPMJPR20A2. R.C.V. was supported by JSPS No P19368. S.O. and M.M.

were supported by JSPS KAKENHI Grant Numbers JP21H05233, JP21H05232, and JP21K14484. K.W. and T.T. acknowledge support from the JSPS KAKENHI (Grant Numbers 20H00354, 21H05233 and 23H02052) and World Premier International Research Center Initiative (WPI), MEXT, Japan.

**Author information**

Authors and Affiliations

**Department of Chemistry, Nagoya University, Nagoya, Aichi, Japan**

Nanami Ichinose, Takato Hotta, Ruben Canton-Vitoria, Fanyu Zeng, Feng Zhang, and Ryo Kitaura

**Innovative Functional Materials Research Institute, National Institute of Advanced Industrial Science and Technology (AIST), Nagoya, Aichi 463-8560, Japan**

Zheng Liu

**Department of Physics, Graduate School of Pure and Applied Sciences, University of Tsukuba, 1-1-1 Tennodai, Tsukuba, Tsukuba 305-8571, Japan**

Mina Maruyama and Susumu Okada

**Research Center for Materials Nanoarchitectonics, National Institute for Materials Science, 1-1 Namiki, Tsukuba 305-0044, Japan**

Takashi Taniguchi and Ryo Kitaura

**Research Center for Functional Materials, National Institute for Materials Science, 1-1 Namiki, Tsukuba 305-0044, Japan**

Kenji Watanabe

Contributions

R.K. designed the experimental study. M.M. led theoretical analyses. S.O. supported the theoretical work. N.I. carried out the MOCVD experiments and spectroscopic characterizations with support from T.H., F.Z., and R.V. T.T. and K.W. prepared hBN crystals. Z.L. carried out STEM-HAADF observations. R.K. wrote the paper with input from other authors.

Corresponding author

Ryo Kitaura and Mina Maruyama

**Ethics declarations**

Competing interests

The authors declare no competing financial interest.

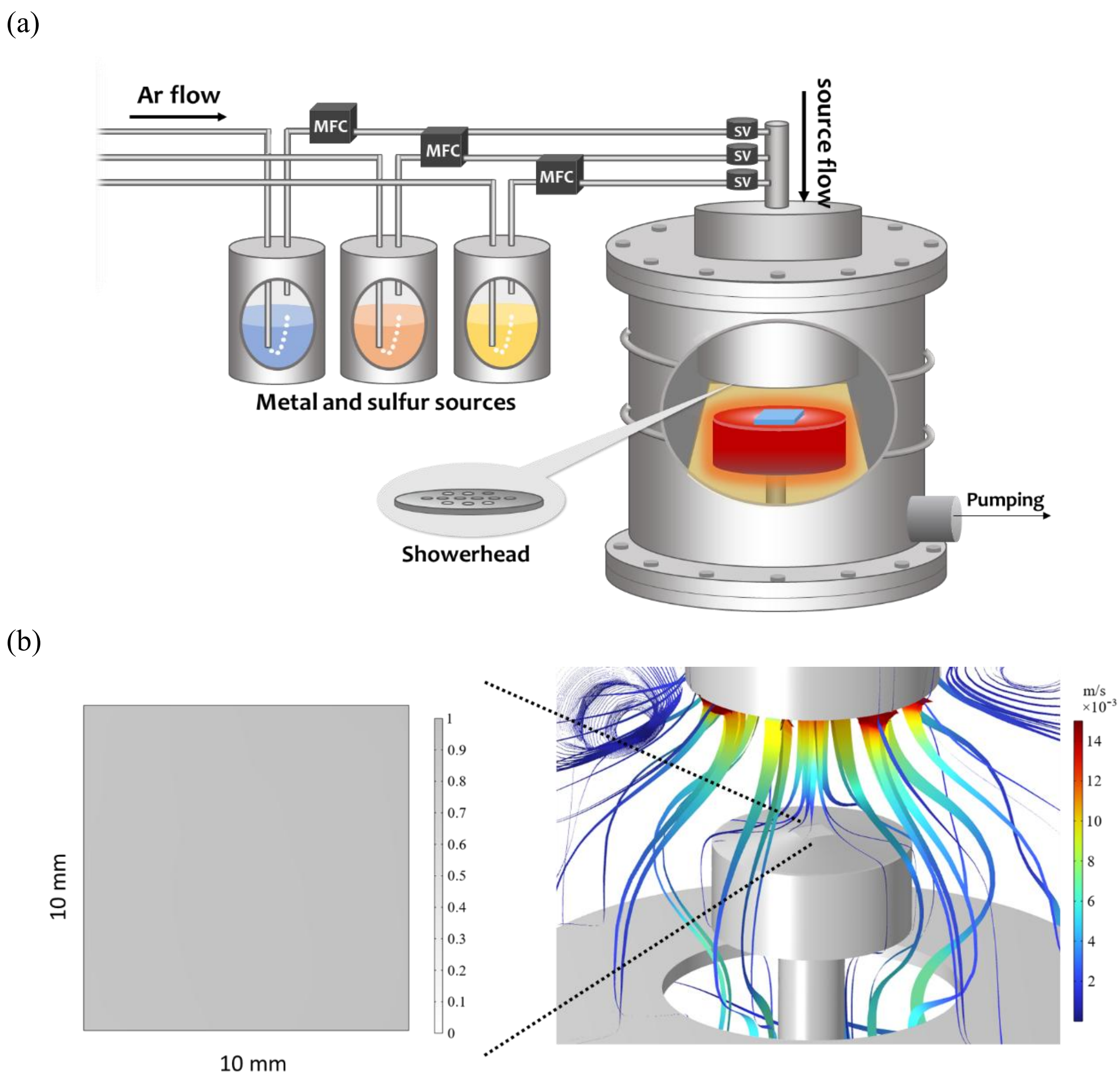

Figure 1. (a) A schematic drawing of the MOCVD chamber for the growth of 1-2mD structures. MFC and SV represent mass flow controllers and solenoid valves, respectively. Metal and sulfur sources are supplied from the top of the chamber by Ar buffer gas flow. The load-lock chamber is omitted for simplicity. (b) Simulated source flow around the substrate (left) and the concentration distribution on the substrate (left). The showerhead locates 5 cm above the substrate/susceptor. Under the susceptor, we put a rectifying plate to make the source flow straight towards the substrate.

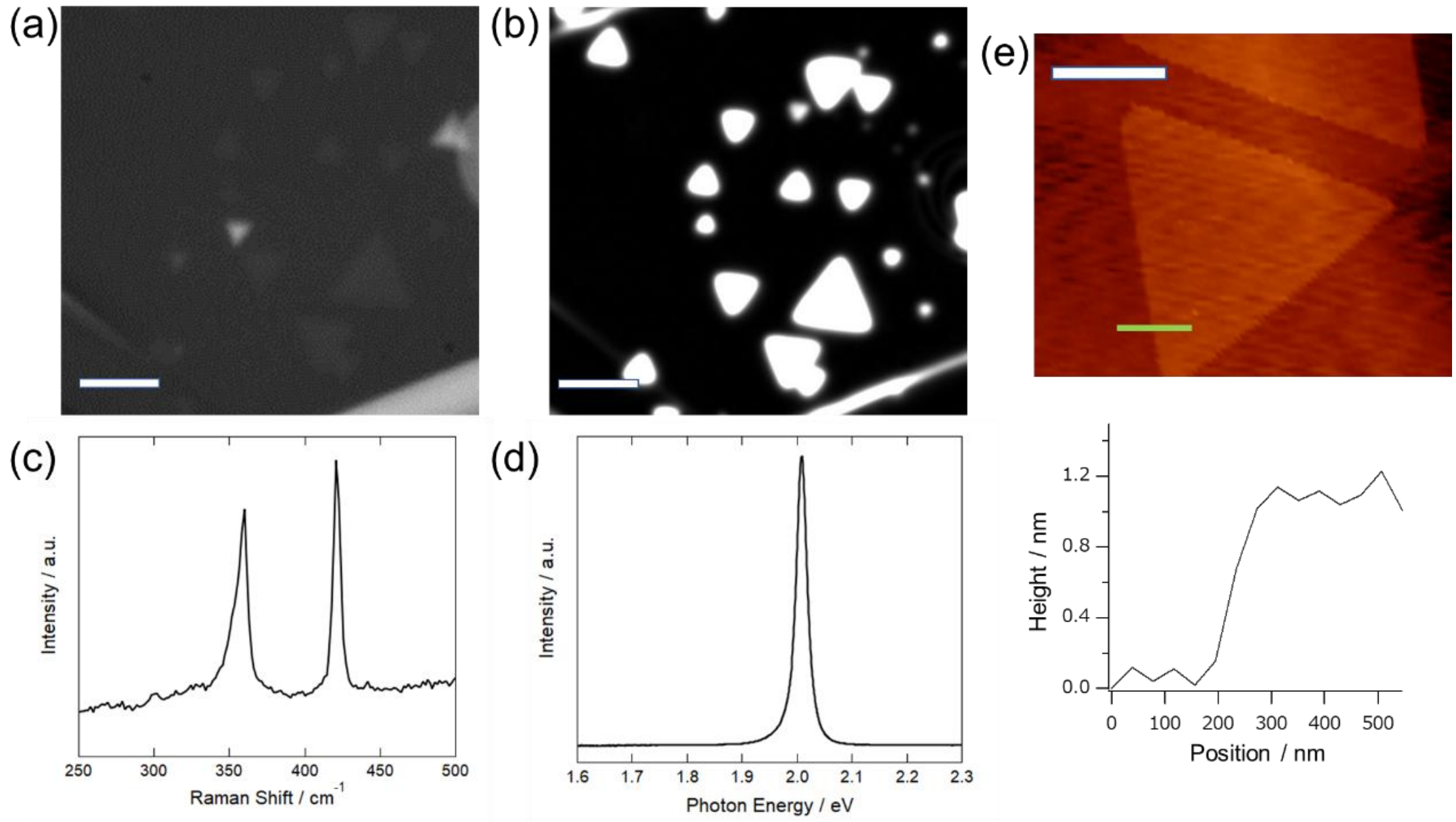

Figure 2. (a) and (b) Optical and PL images of $WS_2$ grown on an hBN flake. Both scale bars correspond to 5 μm. (c) and (d) Typical Raman and PL spectra of $WS_2$ grown on an hBN flake measured at room temperature. Excitation wavelengths of 488 and 532 nm were used for Raman and PL spectra measurements. (e) A typical AFM image and a line profile along the green line in the image. The scale bar corresponds to 1 μm.

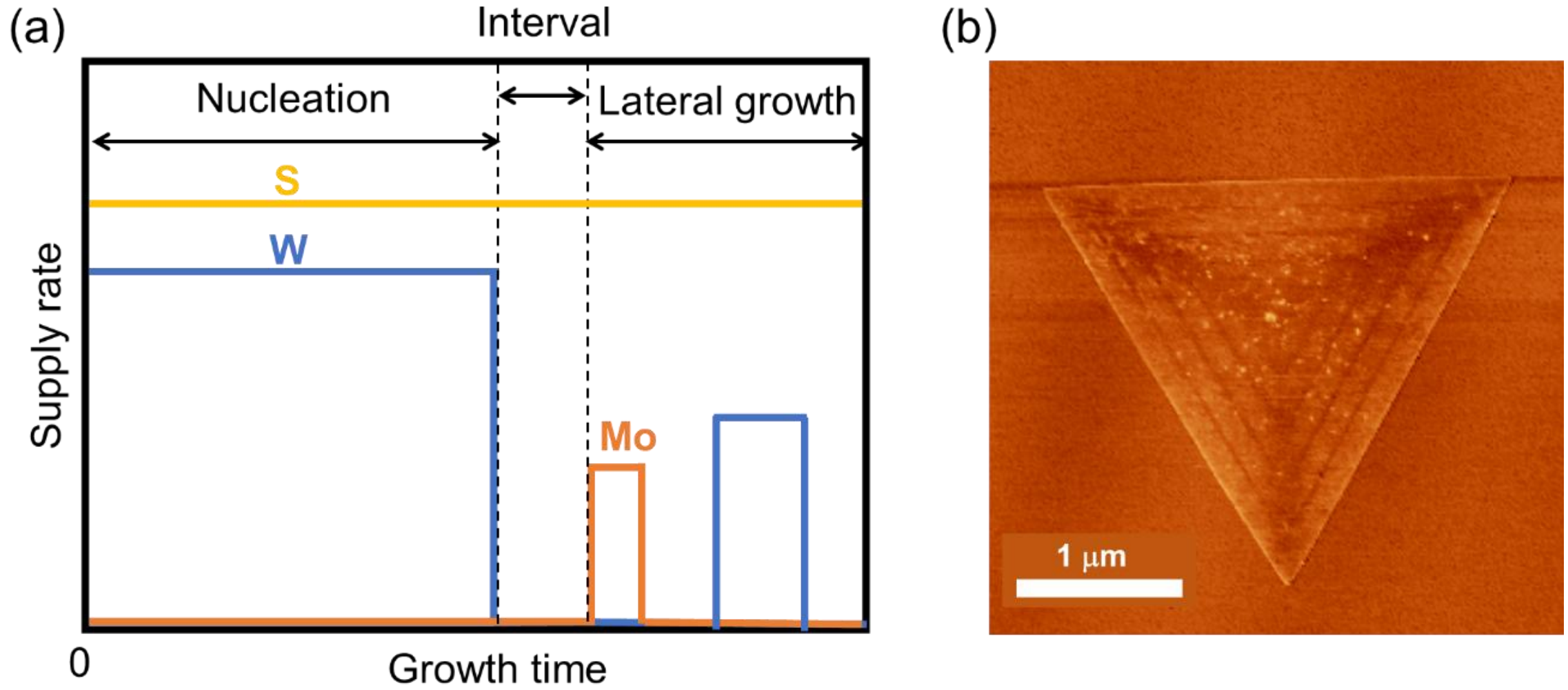

Figure 3. (a) A schematic drawing of source supply for growth of 1-2mD structures composed of $MoS_2$ and $WS_2$. $WS_2$ is grown first, and Mo and W sources are alternatively supplied to form 1-2mD structures. The lateral growth process shown in the Figure is repeated several times to form multiple junctions. The typical supply rate for $WS_2$ nucleation is 80~120 sccm (buffer gas flow) under $W(TBI)_2(DMA)_2$ vapor pressure of 9 mTorr. For lateral growth, we typically used 80 sccm and 30~50 sccm under vapor pressure of 9 mTorr ($W(TBI)_2(DMA)_2$) and 2 mTorr ($Mo(TBI)_2(DMA)_2$), respectively. (b) A typical AFM phase image of a $WS_2/MoS_2$ 1-2mD sample. Four thin ribbons of $MoS_2$ can be seen as dark stripes. The width of the $MoS_2$ ribbons evaluated from this image is less than 10 nm.

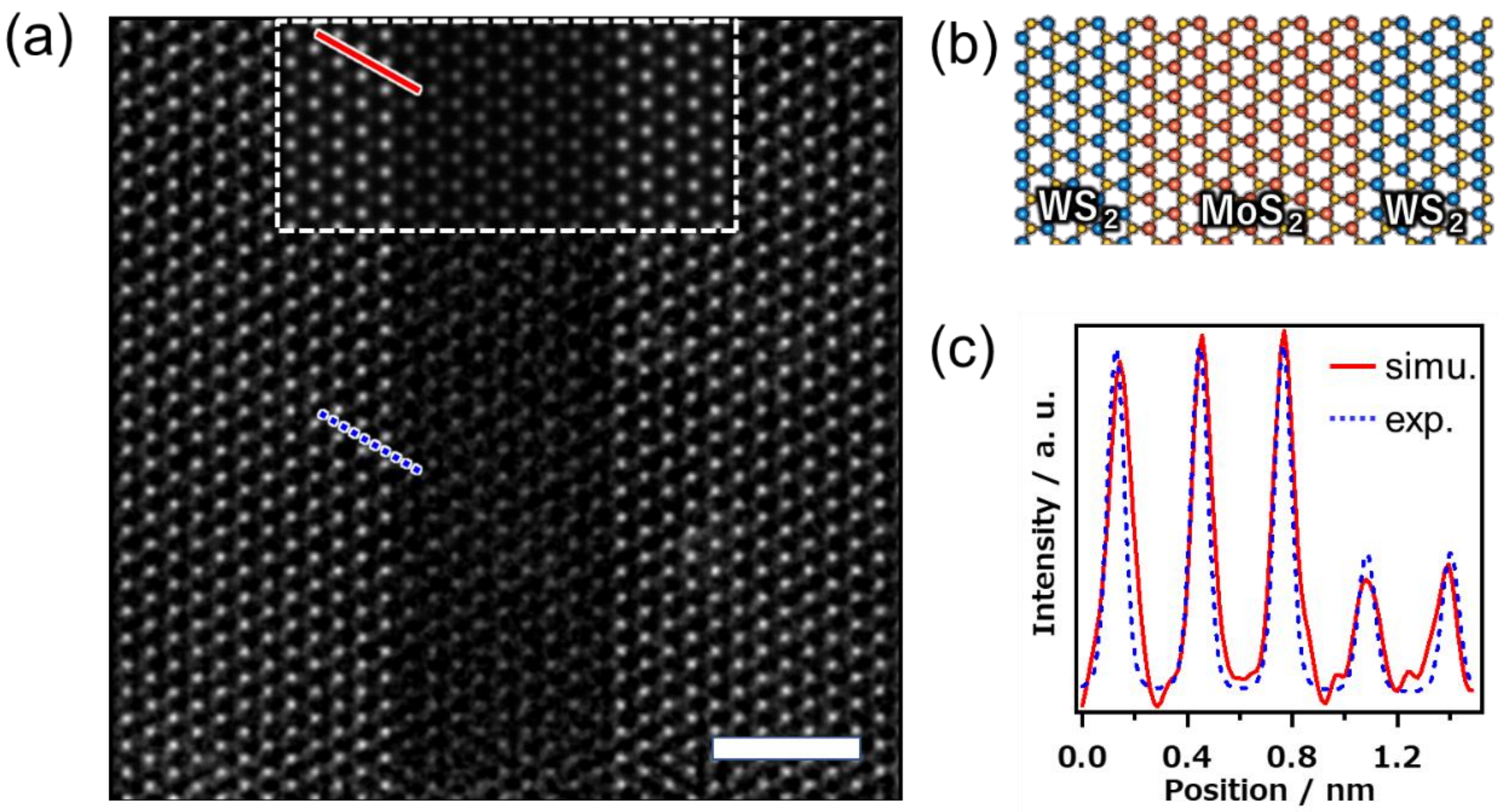

Figure 4. (a) a HAADF-STEM image of a $WS_2/MoS_2$ 1-2mD structure around a junction between $WS_2$ and $MoS_2$. The region with the white-dotted line frame shows a simulated HAADF-STEM image. For the image simulation, we used the multi-slice method with winHREM software. The scale bar corresponds to 2 nm. (b) the corresponding structure model for image simulation. (c) Line profiles of the image contrast along red and blue lines in (a).

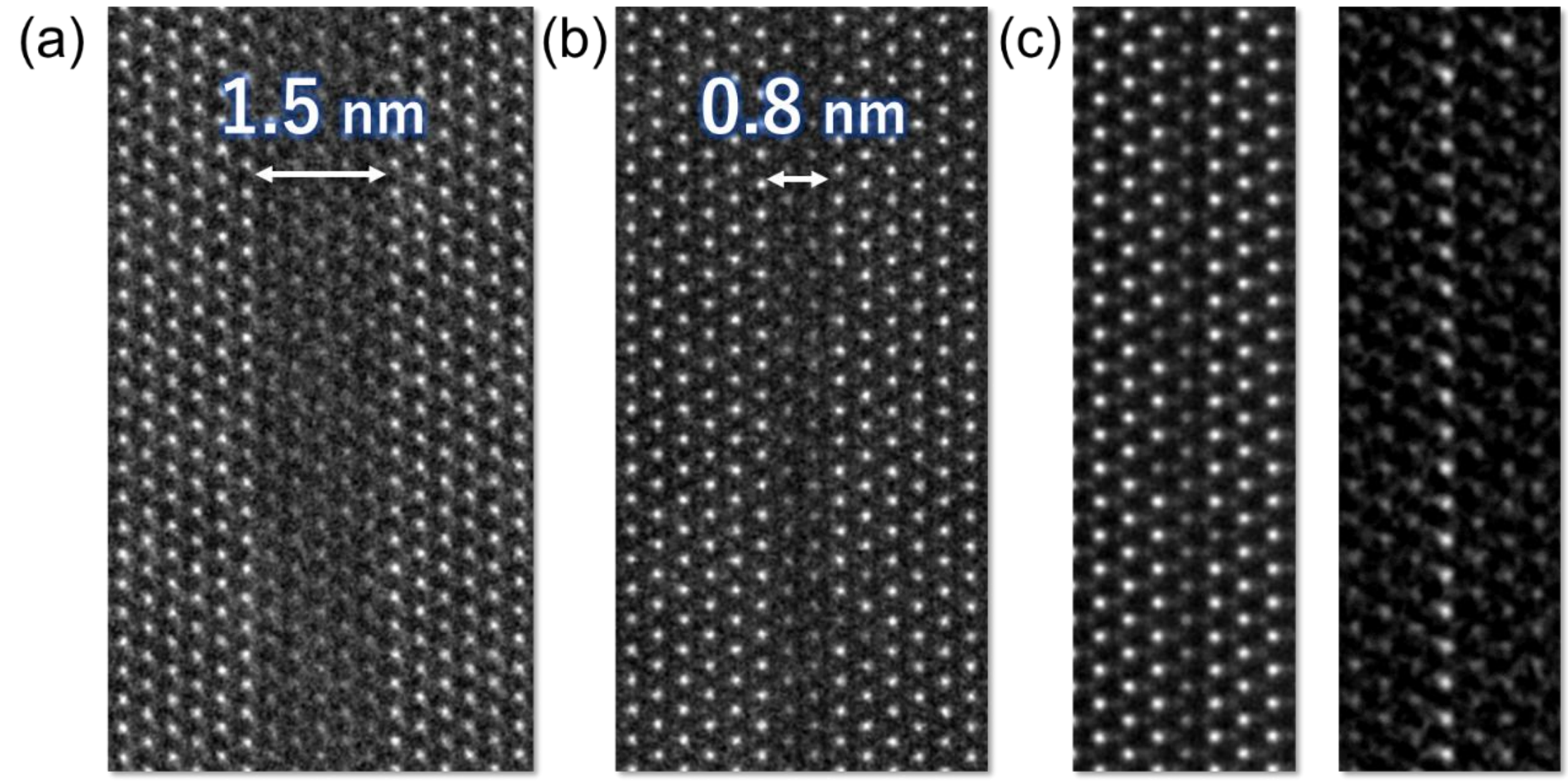

Figure 5. (a), (b), and (c) shows HAADF-STEM images of $WS_2/MoS_2$ 1-2mD structure, whose stripe width is 1.5, 0.8, and single atomic, respectively.

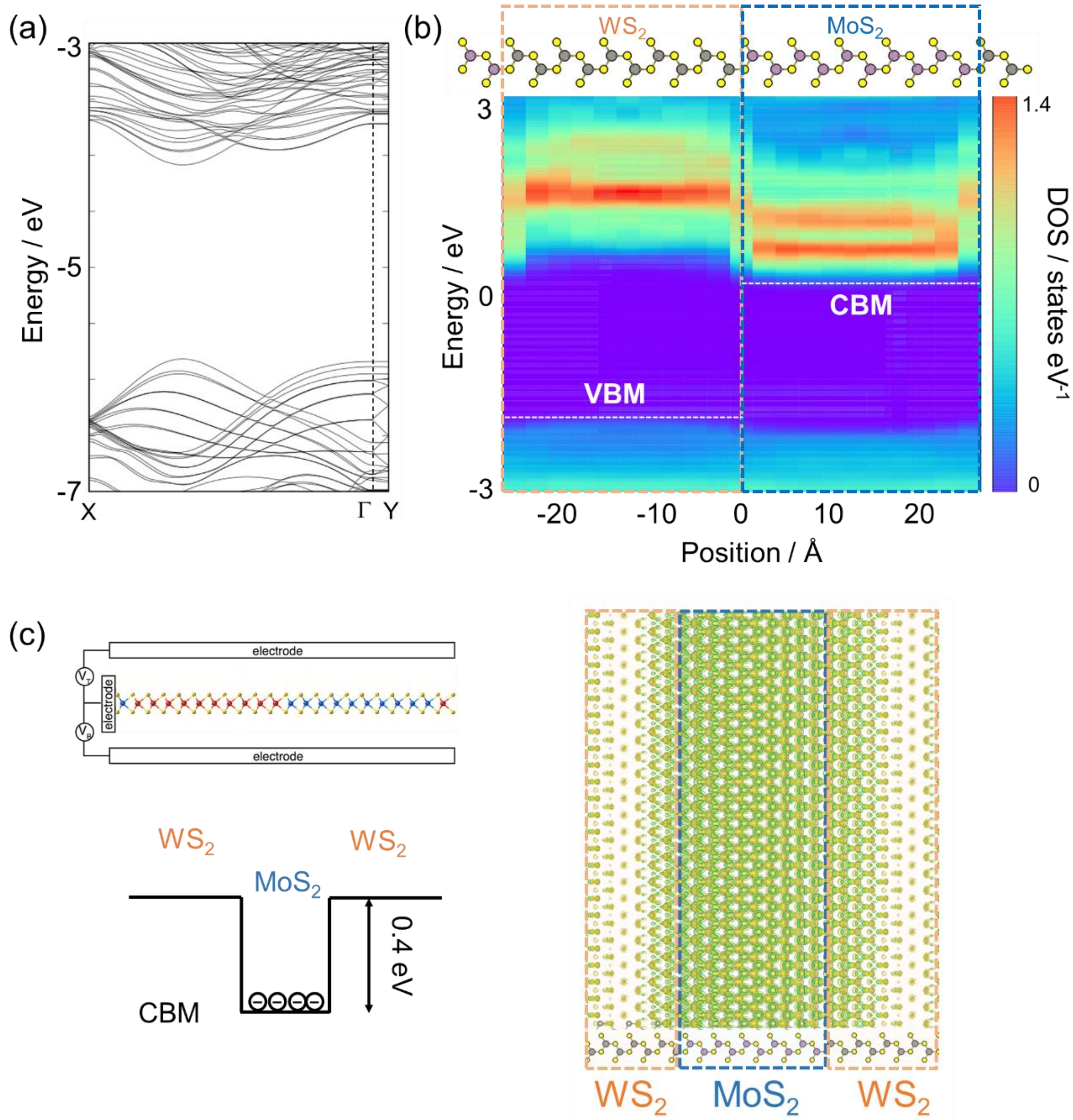

Figure 6. (a) and (b) are electronic band structures and local density of states of a 1-2mD structure of $WS_2/MoS_2$, whose periodicity is 2.45 nm. (c) (upper left) a schematic representation of the dual-gate field-effect transistor used in carrier accumulation simulation. (lower left) an energy diagram around CBM with 1D electron accumulation at the $MoS_2$ stripe. (right) isofurface of electron density at electron density of $10^{13}$ e/cm$^2$.

## REFERENCES

1 Novoselov, K. S. *et al.* Two-dimensional gas of massless Dirac fermions in graphene. *Nature* **438**, 197-200, doi:10.1038/nature04233 (2005).

2 Zhang, Y. B., Tan, Y. W., Stormer, H. L. & Kim, P. Experimental observation of the quantum Hall effect and Berry's phase in graphene. *Nature* **438**, 201-204, doi:10.1038/nature04235 (2005).

3 Castro Neto, A. H., Guinea, F., Peres, N. M. R., Novoselov, K. S. & Geim, A. K. The electronic properties of graphene. *Rev Mod Phys* **81**, 109-162, doi:10.1103/RevModPhys.81.109 (2009).

4 Mak, K. F. *et al.* Measurement of the Optical Conductivity of Graphene. *Phys Rev Lett* **101**, doi:ARTN 196405 10.1103/PhysRevLett.101.196405 (2008).

5 Xiao, D., Liu, G. B., Feng, W. X., Xu, X. D. & Yao, W. Coupled Spin and Valley Physics in Monolayers of MoS2 and Other Group-VI Dichalcogenides. *Phys Rev Lett* **108**, doi:ARTN 196802 10.1103/PhysRevLett.108.196802 (2012).

6 Mak, K. F., Lee, C., Hone, J., Shan, J. & Heinz, T. F. Atomically Thin MoS2: A New Direct-Gap Semiconductor. *Phys Rev Lett* **105**, doi:ARTN 136805 10.1103/PhysRevLett.105.136805 (2010).

7 Mak, K. F., He, K. L., Shan, J. & Heinz, T. F. Control of valley polarization in monolayer MoS2 by optical helicity. *Nat Nanotechnol* **7**, 494-498, doi:10.1038/Nnano.2012.96 (2012).

8 Zeng, H. L., Dai, J. F., Yao, W., Xiao, D. & Cui, X. D. Valley polarization in MoS2 monolayers by optical pumping. *Nat Nanotechnol* **7**, 490-493, doi:10.1038/Nnano.2012.95 (2012).

9 Okada, M. *et al.* Direct and Indirect Interlayer Excitons in a van der Waals Heterostructure of hBN/$WS_2$/$MoS_2$/hBN. *Acs Nano* **12**, 2498-2505, doi:10.1021/acsnano.7b08253 (2018).

10 Mak, K. F. *et al.* Tightly bound trions in monolayer MoS2. *Nat Mater* **12**, 207-211, doi:10.1038/Nmat3505 (2013).

11 Novoselov, K. S. *et al.* Two-dimensional atomic crystals. *P Natl Acad Sci USA* **102**, 10451-10453, doi:10.1073/pnas.0502848102 (2005).

12 Qian, X. F., Liu, J. W., Fu, L. & Li, J. Quantum spin Hall effect in two-dimensional transition metal dichalcogenides. *Science* **346**, 1344-1347, doi:10.1126/science.1256815 (2014).

13 Huang, B. *et al.* Layer-dependent ferromagnetism in a van der Waals crystal down to the monolayer limit. *Nature* **546**, 270-273, doi:10.1038/nature22391 (2017).

14 Avalos-Ovando, O., Mastrogiuseppe, D. & Ulloa, S. E. Lateral heterostructures and one-dimensional interfaces in 2D transition metal dichalcogenides. *J Phys-Condens Mat* **31**, doi:ARTN 213001 10.1088/1361-648X/ab0970 (2019).

15 Levine, B. F. Quantum-Well Infrared Photodetectors. *J Appl Phys* **74**, R1-R81, doi:Doi 10.1063/1.354252 (1993).

16 Arakawa, Y. & Yariv, A. Quantum-Well Lasers Gain, Spectra, Dynamics. *Ieee J Quantum Elect* **22**, 1887-1899, doi:Doi 10.1109/Jqe.1986.1073185 (1986).

17 Gong, Y. J. *et al.* Vertical and in-plane heterostructures from $WS_2$/$MoS_2$ monolayers. *Nat Mater*

**13**, 1135-1142, doi:10.1038/Nmat4091 (2014).

18 Zhang, Z. W. *et al.* Robust epitaxial growth of two-dimensional heterostructures, multiheterostructures, and superlattices. *Science* **357**, 788-+, doi:DOI 10.1126/science.aan6814 (2017).

19 Duan, X. D. *et al.* Lateral epitaxial growth of two-dimensional layered semiconductor heterojunctions. *Nat Nanotechnol* **9**, 1024-1030, doi:10.1038/Nnano.2014.222 (2014).

20 Lee, J. *et al.* Direct Epitaxial Synthesis of Selective Two-Dimensional Lateral Heterostructures. *Acs Nano* **13**, 13047-13055, doi:10.1021/acsnano.9b05722 (2019).

21 Kobayashi, Y. *et al.* Continuous Heteroepitaxy of Two Dimensional Heterostructures Based on Layered Chalcogenides. *Acs Nano* **13**, 7527-7535, doi:10.1021/acsnano.8b07991 (2019).

22 Xie, S. E. *et al.* Coherent, atomically thin transition-metal dichalcogenide superlattices with engineered strain. *Science* **359**, 1131-1135, doi:10.1126/science.aao5360 (2018).

23 Zhou, W. *et al.* Dislocation-driven growth of two-dimensional lateral quantum-well superlattices. *Sci Adv* **4**, doi:ARTN eaap9096 10.1126/sciadv.aap9096 (2018).

24 Han, Y. *et al.* Sub-nanometre channels embedded in two-dimensional materials. *Nat Mater* **17**, 129-+, doi:10.1038/nmat5038 (2018).

25 Kitaura, R. *et al.* Chemical Vapor Deposition Growth of Graphene and Related Materials. *J Phys Soc Jpn* **84**, doi:Artn 121013 10.7566/Jpsj.84.121013 (2015).

26 Yang, P. F. *et al.* Batch production of 6-inch uniform monolayer molybdenum disulfide catalyzed by sodium in glass. *Nat Commun* **9**, doi:ARTN 979 10.1038/s41467-018-03388-5 (2018).

27 Li, S. S. *et al.* Halide-assisted atmospheric pressure growth of large $WSe_2$ and $WS_2$ monolayer crystals. *Appl Mater Today* **1**, 60-66, doi:10.1016/j.apmt.2015.09.001 (2015).

28 Okada, M. *et al.* Direct Chemical Vapor Deposition Growth of $WS_2$ Atomic Layers on Hexagonal Boron Nitride. *Acs Nano* **8**, 8273-8277, doi:10.1021/nn503093k (2014).

29 Hotta, T. *et al.* Enhanced Exciton-Exciton Collisions in an Ultraflat Monolayer MoSe2 Prepared through Deterministic Flattening. *Acs Nano* **15**, 1370-1377, doi:10.1021/acsnano.0c08642 (2021).

30 Okada, M. *et al.* Observation of biexcitonic emission at extremely low power density in tungsten disulfide atomic layers grown on hexagonal boron nitride. *Sci Rep.* **7**, doi:ARTN 322 10.1038/s41598-017-00068-0 (2017).

31 Ishizuka, K. A practical approach for STEM image simulation based on the FFT multislice method. *Ultramicroscopy* **90**, 71-83, doi:Pii S0304-3991(01)00145-0 Doi 10.1016/S0304-3991(01)00145-0 (2002).

32 Otani, M. & Sugino, O. First-principles calculations of charged surfaces and interfaces: A plane-wave nonrepeated slab approach. *Phys Rev B* **73**, doi:ARTN 115407 10.1103/PhysRevB.73.115407 (2006).

# Supplementary materials

## Two-dimensional atomic-scale ultrathin lateral heterostructures

Nanami Ichinose[1], Mina Maruyama[5*], Takato Hotta[1], Zheng Liu[3], Ruben Canton-Vitoria[1], Susumu Okada[5], Fanyu Zeng[1], Feng Zhang[1], Takashi Taniguchi[3], Kenji Watanabe[4], and Ryo Kitaura[3,*]

[1] *Department of Chemistry, Nagoya University, Nagoya, Aichi 464-8602 Japan*

[2] *Innovative Functional Materials Research Institute, National Institute of Advanced Industrial Science and Technology (AIST), Nagoya, Aichi 463-8560, Japan*

[3] *Research Center for Materials Nanoarchitectonics, National Institute for Materials Science, 1-1 Namiki, Tsukuba 305-0044, Japan*

[4] *Research Center for Functional Materials, National Institute for Materials Science, 1-1 Namiki, Tsukuba 305-0044, Japan*

[5] *Department of Physics, Graduate School of Pure and Applied Sciences, University of Tsukuba, 1-1-1 Tennodai, Tsukuba, Tsukuba 305-8571, Japan*

Corresponding Author: R. Kitaura, KITAURA.Ryo@nims.go.jp, r.kitaura@nagoya-u.jp; M. Maruyama, mmaruyama@comas-tsukuba.jp

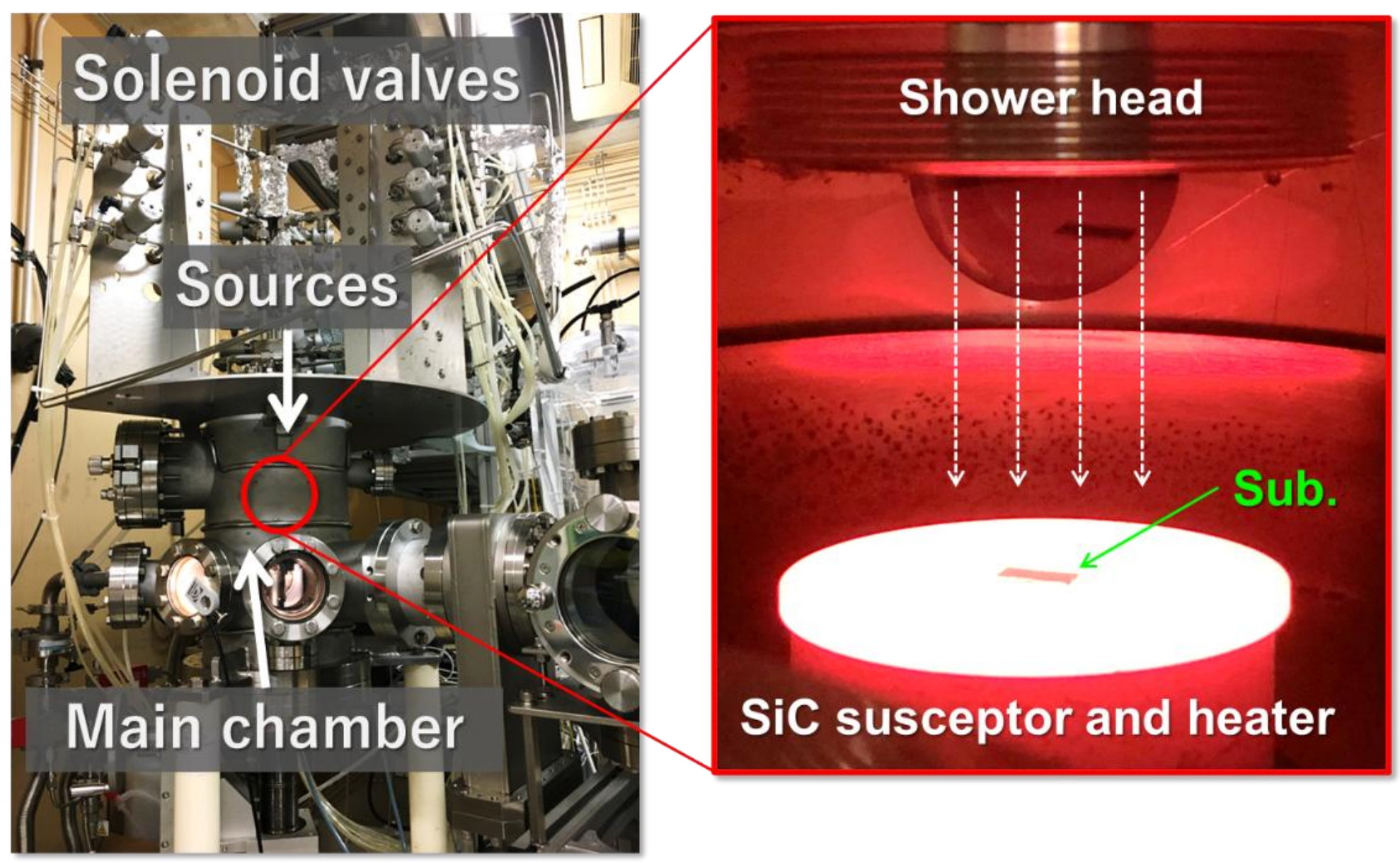

Figure S1 Pictures of the MOCVD chamber and the inside of the chamber during a growth process.

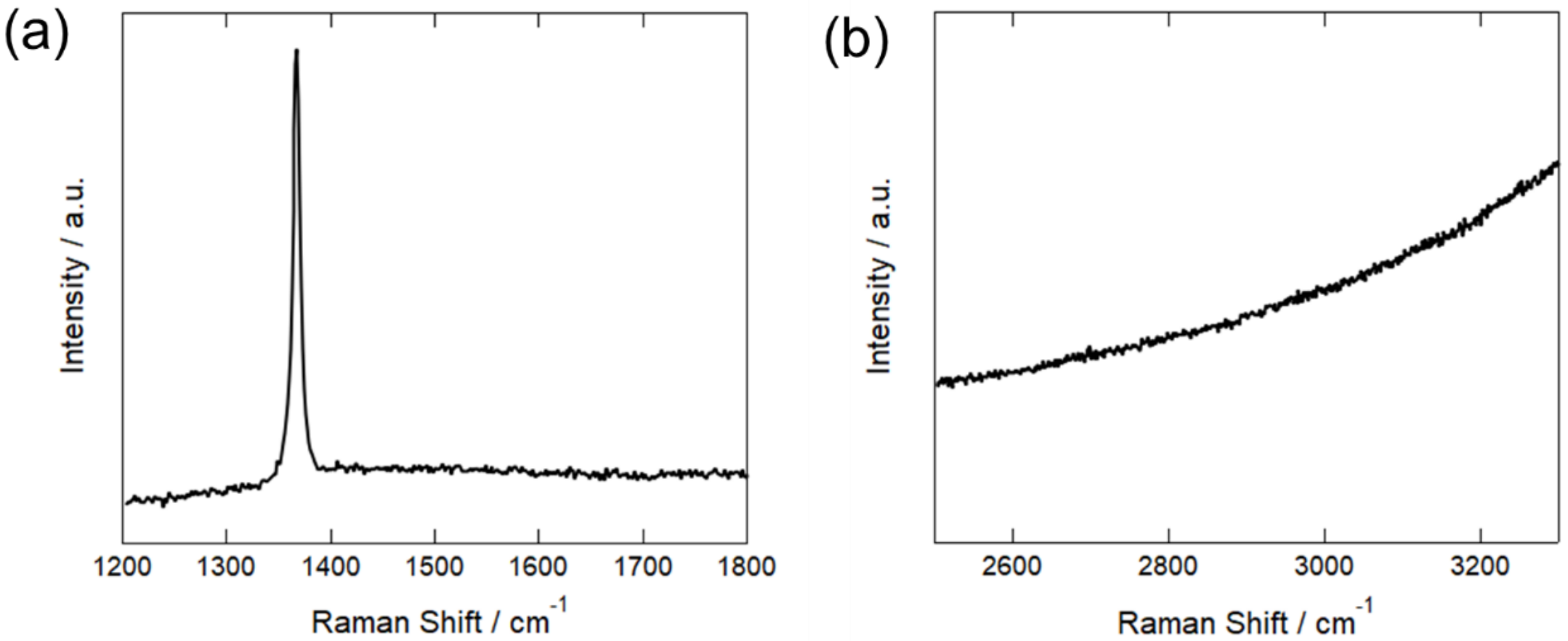

Figure S2 (a), (b) Raman spectra of $WS_2$ grown on an hBN flake measured at room temperature with an excitation wavelength of 488 nm.

Figures show typical Raman spectra of a single crystal of $WS_2$ grown on an hBN flake. The strong peak at 1358 $cm^{-1}$ in the left spectrum is assigned to the $E_{2g}$ mode of hBN. Except for hBN $E_{2g}$, no pronounced Raman band exists in 1200 – 1800 and 2500 – 3300 $cm^{-1}$, meaning that organic moieties of MOCVD sources do not remain in the final product.

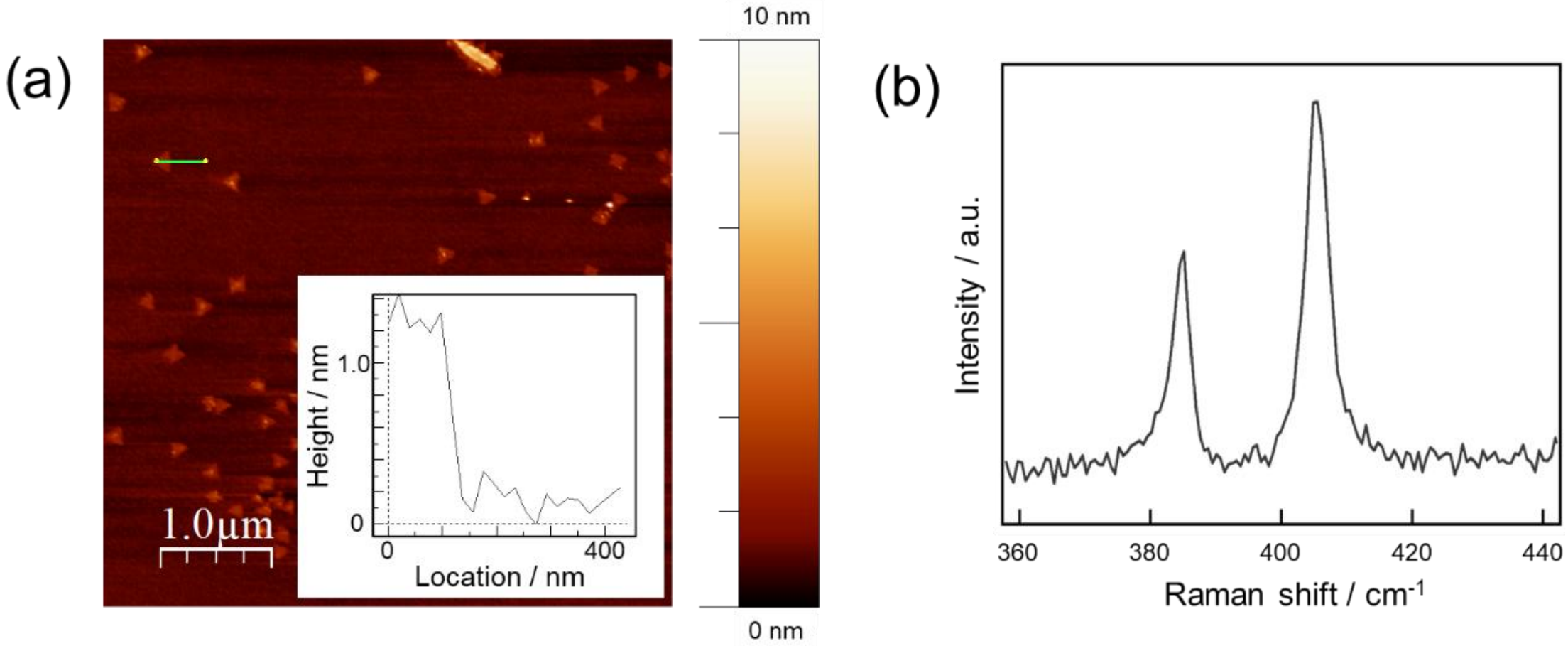

Figures S3 (a), (b) a typical AFM image and a Raman spectrum of $MoS_2$ grown on an hBN flake. The Raman spectrum was measured at room temperature with an excitation wavelength of 488 nm.

Triangular contrasts, whose height is consistent with monolayer $MoS_2$, exist over the flake. The observed two Raman bands are characteristic of $MoS_2$.

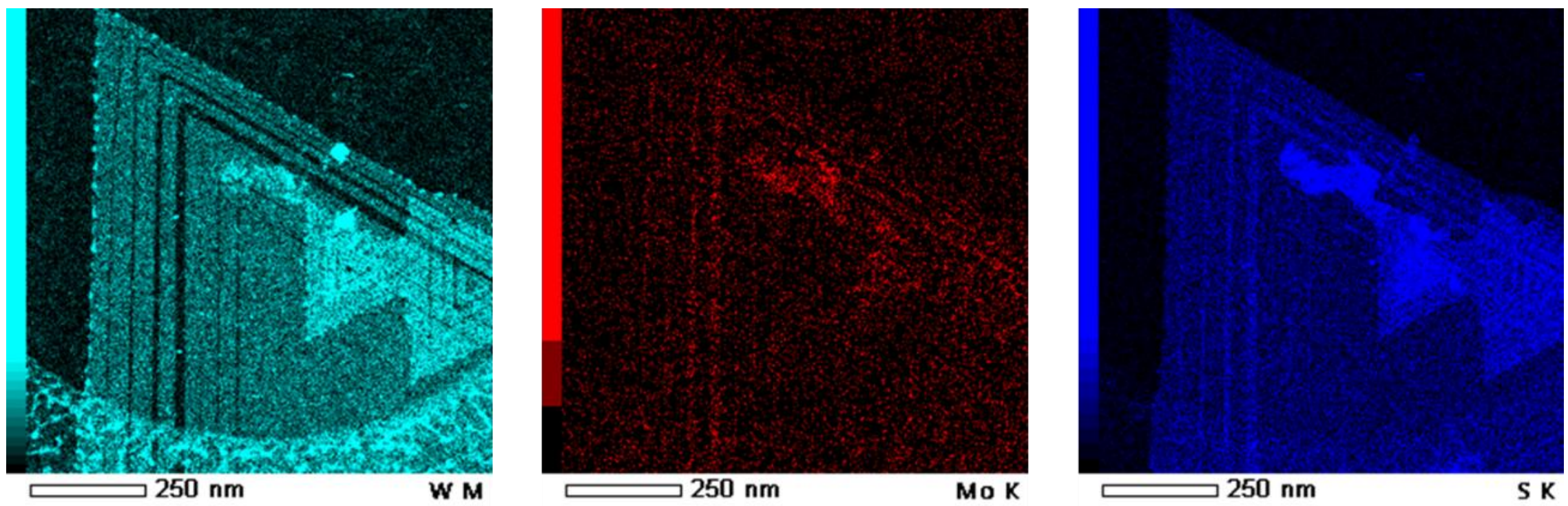

Figure S4 Energy dispersive X-ray analysis maps of a 1-2mD structure of $MoS_2$ and $WS_2$. (left) W M, (middle) Mo K, and (right) S K regions.

The figure shows that Sulfur signals exist over the sample, whereas W and Mo signals localize at each stripe. This EDX map is consistent with the formation of a 1-2mD structure of $MoS_2$ and $WS_2$.

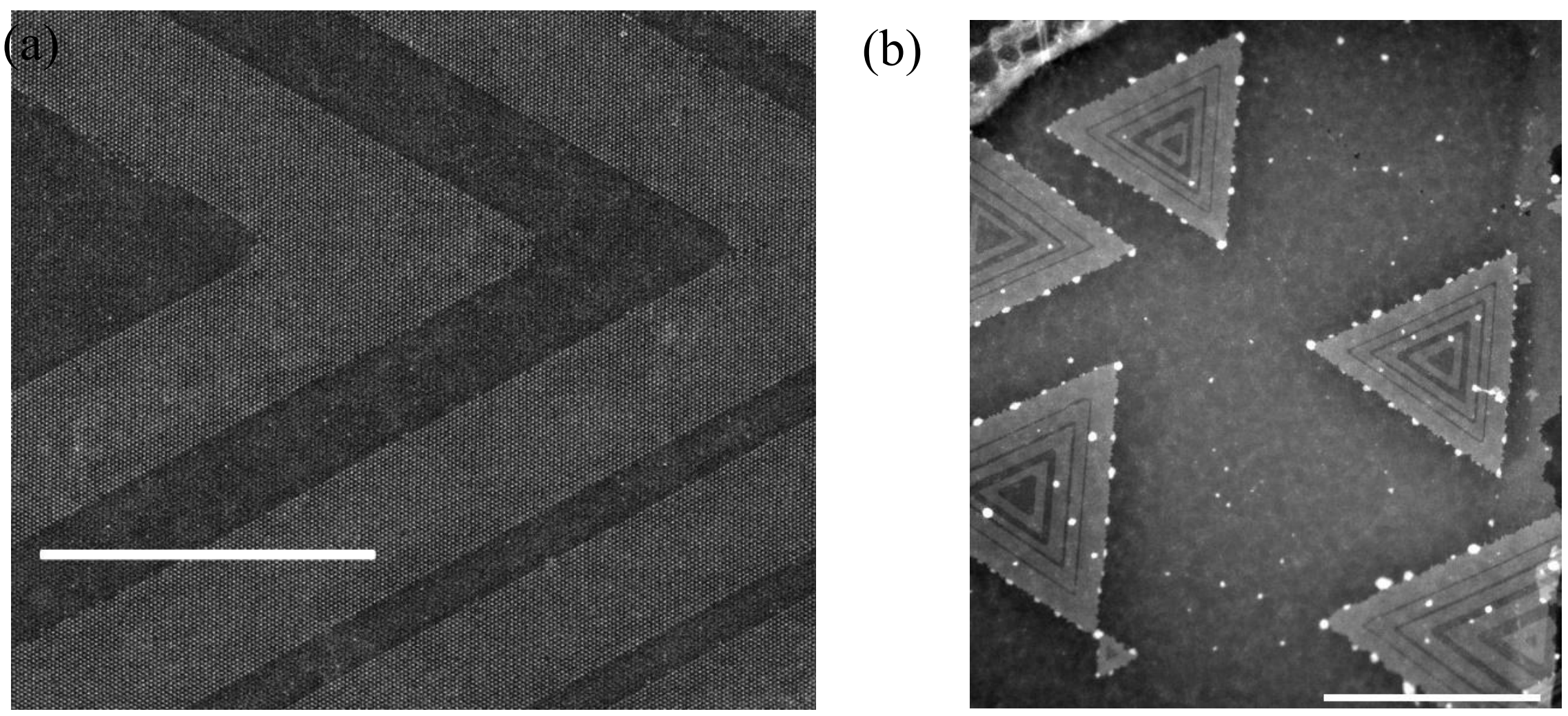

Figure S5 Low-magnification HAADF-STEM images of 1-2mD structures composed of $MoS_2$ and $WS_2$. Scale bars in (a) and (b) correspond to 25 and 200 nm, respectively.

As seen in the images, the number of bright spots, which correspond to W atoms, in dark stripes ($MoS_2$ stripes) is small. These images strongly suggest that the fast switchings of source supply enabled by solenoid valve operations during MOCVD growth processes minimize the mixing of W and Mo in products.

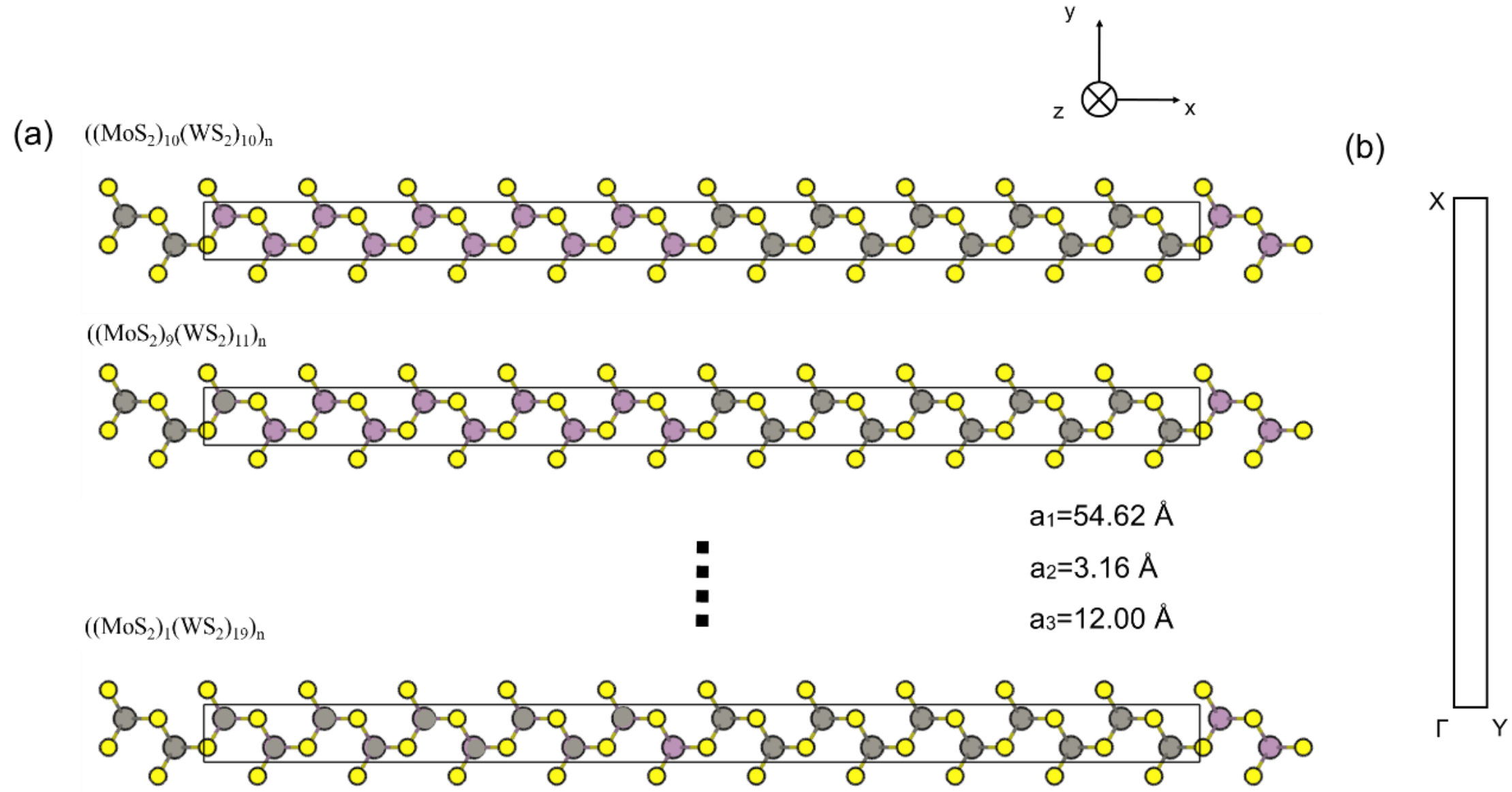

Figure S6 Unit cells and the corresponding Brillouin zone of $MoS_2/WS_2$ lateral heterostructures with different stripe widths. Purple, gray, and yellow circles represent Mo, W, and S atoms, respectively.

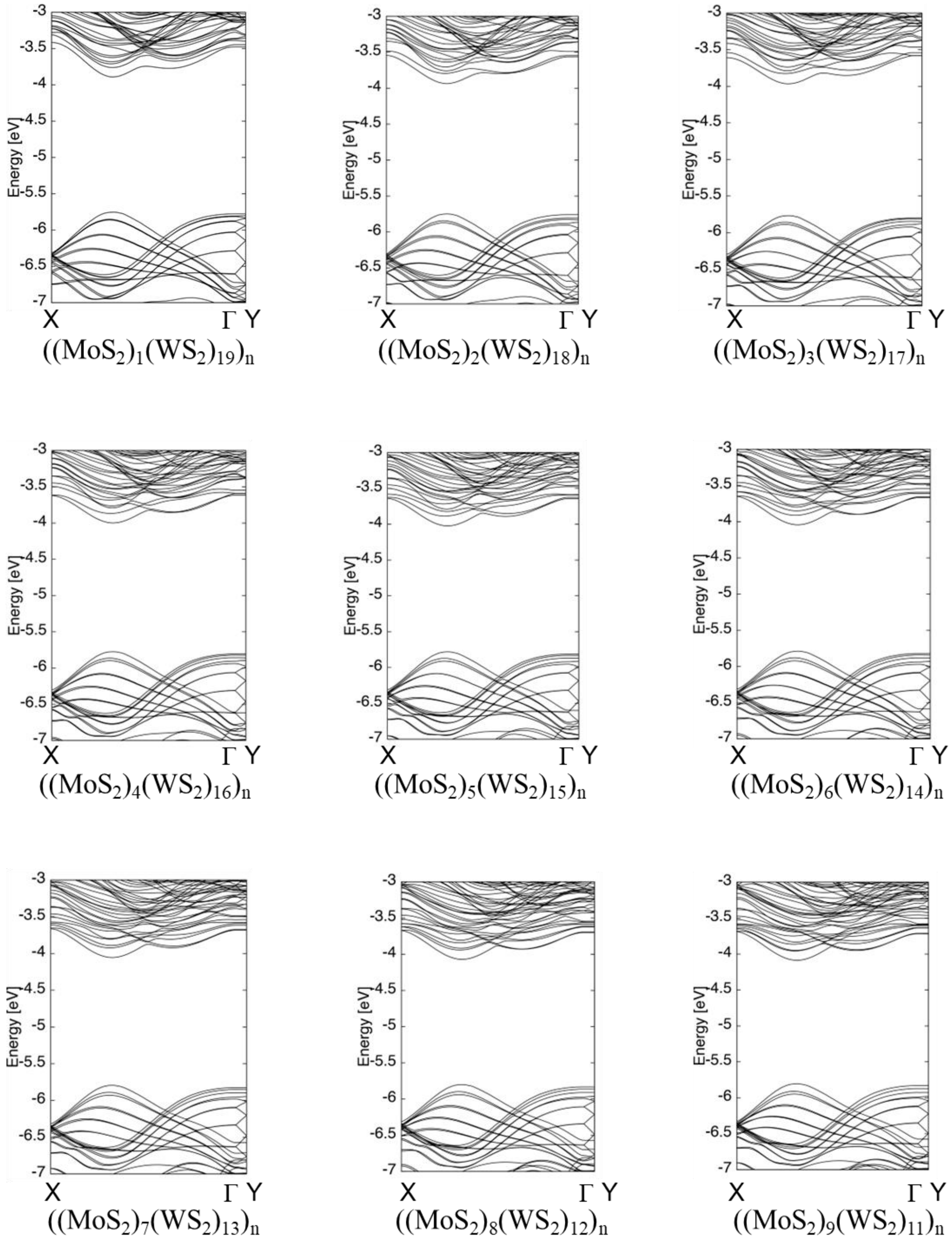

Figure S7 Electronic energy band of $MoS_2$/$WS_2$ lateral heterostructures with different stripe widths. In all cases, the total number of metal atoms (W and Mo) in a unit cell is the same. The energies are measured from the vacuum level in all cases.

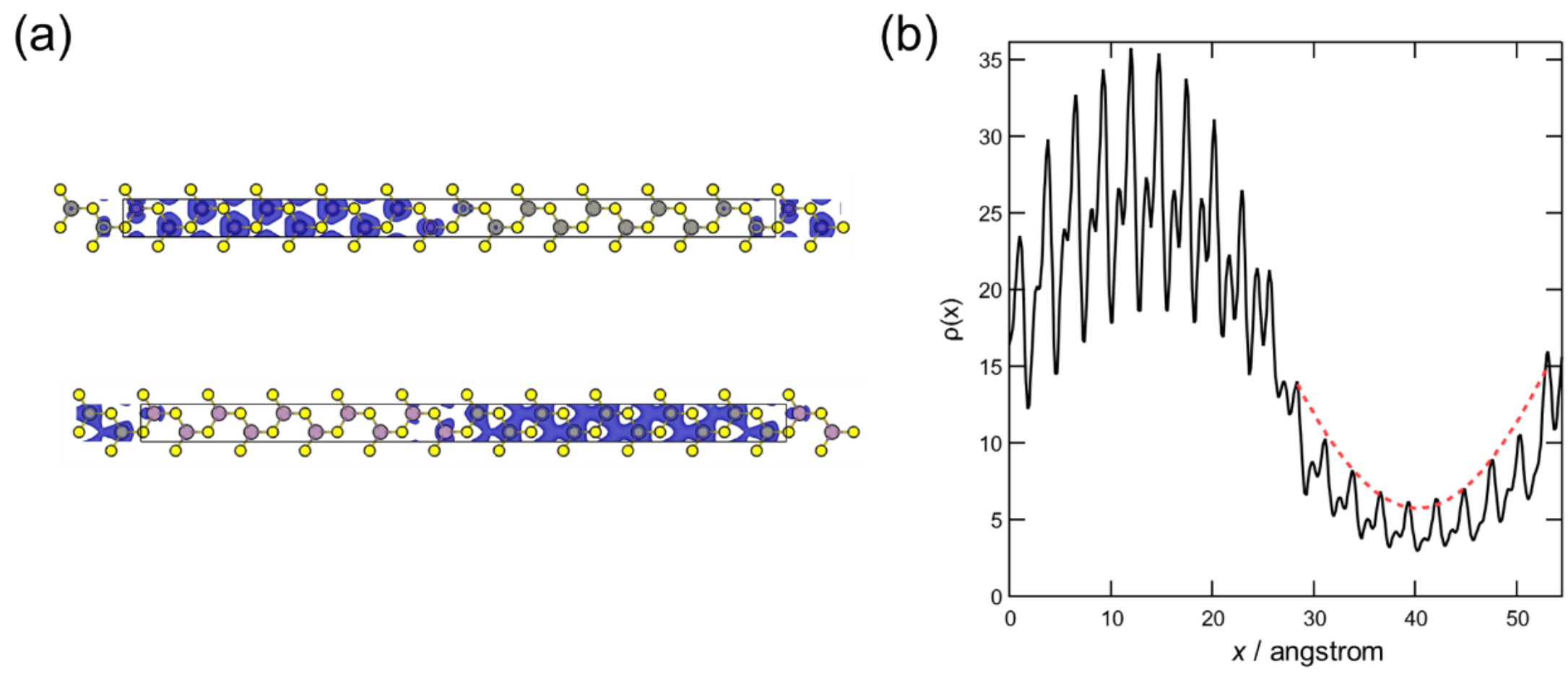

Figure S8 (a) Isosurface of wave functions at the CBM (upper) and VBM (lower) (b) Squared wave function integrated on the planes normal to the striped structure. The horizontal axis corresponds to the location along the direction perpendicular to the 1D stripe axis. The red dotted line corresponds to a fitting curve assuming exponential decay.

The figure shows a squared wave function of the CBM integrated on the planes normal to the striped structure: $\rho(x) = \int dy\, dz\, |\Phi_{CBM}|^2$, where the $\Phi_{CBM}$ is the wave function of the CBM, y, and z correspond to the direction along the border and normal to the plane. $\rho(x)$ shows a long-period wave profile (envelope curve) in addition to a sawtooth pattern arising from the atomic structure of $MoS_2/WS_2$. We fitted the envelope curve assuming exponential decay of wavefunction, extracting a penetration length of 1.6 nm.

## Supplementary discussion 1: carrier injection into 1-2mD structures of $MoS_2$ and $WS_2$:

The energy offset at CBM is ca. 0.4 eV in 1-2mD structures of $MoS_2$ and $WS_2$. To induce 1D to 2D dimensionality alternation, we need to fill 1D $MoS_2$ states below the bottom of the $WS_2$ state. Usually, shifting the Fermi level by 0.4 eV is challenging, but it is feasible in nano-scale 1D stipes, which can be realized in the present MOCVD method, because the number of states of 1D $MoS_2$ stripes is small when stripe width reaches nano-scale. Due to the small number of states, low-level carrier injection can significantly alter the potential at 1D stripes, as discussed below. In the $((WS_2)_{10}/(MoS_2)_{10})_n$, the effective mass $m^*$ of the CBM band (Mo 1D stripes) is 0.46 $m_0$, which yields the relation between energy $E$ and the number of states $N$ as follows.

$$N/L = 2\frac{\sqrt{2m^*}}{\pi\hbar}\sqrt{E} = 2.2\sqrt{E} \text{ state/nm}$$

, where $L$ corresponds to the length of 1D stripes, and $E$ represents energy from the bottom of CBM. Substituting 0.4 eV to the above equation gives 1.4 states/nm, which corresponds to the number of states of 1D $MoS_2$ below the bottom of the 2D $WS_2$ state. Assuming a 1D stripe embedded in a 1-2mD structure of $1 \times 1$ $\mu m^2$, an electron density of $1.4 \times 10^{11}$ / $cm^2$ can move the potential at 1D stripes 0.4 eV upward, leading to 1D to 2D dimensionality alternation. Furthermore, charrier accumulation into small regions causes energy increase arising from charging. The charging energy can be significant when carrier injection occurs in 1D small objects. For example, single electron charging energy in a carbon nanotube with a length of 300 nm (electron density is only $3.3 \times 10^{-2}$ / nm) is reported to be 6.7 meV [1]. The charging energy under higher electron density can be significant, which compensates for the energy offset at CBM, further facilitating 1D to 2D crossover.

[1] S. Moriyama et al., *Phys. Rev. Lett.*, 94, 186806 (2005)

## Supplementary information: flow simulation details:

We performed flow simulation using the commercial finite element software COMSOL, where simulation modules, such as laminar flow, heat transfer, and mass transfer, are included. The conditions, including the chamber design, growth temperature, and flow rates, are set according to the actual experiments. Because the details of the source decomposition and TMD formation reaction in MOCVD growth are unknown, we did not consider the chemical reaction and analyzed the results using the precursor concentration as an indicator. In the simulation, Navier–Stokes equation,

$$\rho(\boldsymbol{u} \cdot \nabla \boldsymbol{u}) = \nabla \cdot \left[-p\boldsymbol{I} + \mu(\nabla \boldsymbol{u} + (\nabla \boldsymbol{u})^T) - \frac{2}{3}\mu(\nabla \cdot \boldsymbol{u})\boldsymbol{I}\right] + \boldsymbol{F};\ \nabla \cdot (\rho \boldsymbol{u}) = 0$$

,where $\rho$ is the density, $\boldsymbol{u}$ is the velocity field, $p$ is pressure, $\boldsymbol{I}$ is the unit matrix, $\mu$ is the dynamic viscosity, and $\boldsymbol{F}$ is the volumetric applied force, are solved to elucidate the flow and the precursor source concentration in the chamber. The heat transfer equation used is

$$\rho C_p \boldsymbol{u} \cdot \nabla T + \nabla \cdot \boldsymbol{q} = Q; \boldsymbol{q} = -k\nabla T$$

, where $C_p$ is the heat capacity at constant pressure, $\boldsymbol{q}$ is the heat flux vector, $Q$ is the heating source, and $k$ is the thermal conductivity coefficient. In this study, we set the temperature of the object stage as a constant, so the heat source $Q$ is zero in our model.

The mass transfer equation is

$$\nabla \cdot -D_i \nabla c_i + \boldsymbol{u} \cdot \nabla c_i = R_i$$

, where $D$ is the diffusion coefficient, $c$ is the precursor concentration, and $R$ is the source of the precursor.